\documentclass[%
      aps,
      prx, 
      reprint,
      twocolumn,
      showpacs]{revtex4-1}
\usepackage[ colorlinks,
    linkcolor={blue},
    citecolor={blue},
    urlcolor={blue}]{hyperref}
\usepackage{comment}
\usepackage{float}
\usepackage{dsfont}

\newcommand{\be}{\begin{equation}}
\newcommand{\ee}{\end{equation}}
\newcommand{\bea}{\begin{eqnarray}}
\newcommand{\eea}{\end{eqnarray}}
\newcommand{\br}{{\bf r}}

\newcommand{\bj}{{\bf j}}

\newcommand{\un}{\underline{n}}

\newcommand{\intbz}{\int\limits_\text{BZ}}

\newcommand{\intbzdkpi}{ \intbz \frac{\dd\bk}{(2\pi)^d}}

\newcommand{\vF}{v_\text{F}}

\usepackage{graphicx,color,amssymb,amsfonts}
\definecolor{red}{rgb}{1,0,0}
\definecolor{green}{rgb}{0,0.75,0}
\definecolor{pink}{rgb}{0,1,1}
\definecolor{blue}{rgb}{0,0,1}
\definecolor{orange}{rgb}{1,0.4,0}
\usepackage{bm}
\usepackage{amsmath}
\usepackage{amssymb}
\usepackage{amsfonts}
\usepackage{upgreek}
\usepackage{mathbbol}
\usepackage{mathtools} 
\usepackage{hyperref}[urlcolor=blue]
\usepackage{bbold} 
\usepackage[normalem]{ulem}

\usepackage{graphicx}
\usepackage{mwe}
\usepackage{pgfplots}
\pgfplotsset{
compat=1.5,
tick label style={font=\small},
every axis/.append style={line width=0.8pt, tick style={line width=0.7pt}}
} 
\usetikzlibrary{shapes,arrows,fit,calc,positioning,arrows}

\DeclareMathAlphabet{\mathpzc}{OT1}{pzc}{m}{it}
\DeclareMathAlphabet\mathbfcal{OMS}{cmsy}{b}{n}

\newcommand{\ci}{{ i}}

\newcommand{\bE}{{\bf E}}
 
\newcommand{\bA}{{\bf A}}

\newcommand{\bfd}{\mathbf{d}}

\newcommand{\dd}{\text{d}}

\newcommand{\ketnk}{|n\bk\rangle}

\definecolor{darkred}{rgb}{0.66666667,0,0}

\newcommand\eqt{\hspace{0.17em}{=}\hspace{0.17em}}
\newcommand\lltext{\hspace{0.17em}{\ll}\hspace{0.17em}}

\newcommand\cdott{\hspace{0.17em}{\cdot}\hspace{0.17em}}

\newcommand\coloneqqt{\hspace{0.17em}{\coloneqq}\hspace{0.17em}}

\newcommand\pt{\hspace{0.17em}{+}\hspace{0.17em}}
\newcommand\mt{\hspace{0.17em}{-}\hspace{0.17em}}

    \newcommand\timest{\hspace{0.12em}{\times}\hspace{0.12em}}

\newcommand{\bv}{\mathbf{v}}

\newcommand{\jpara}{j_\parallel}
\newcommand{\dcr}{\mathbf{j}^{(\infty)}}

\newcommand{\dcrdomparallel}{\hat{j}_\parallel^{(\infty)}}
\newcommand{\dcrparallel}{j_\parallel^{(\infty)}}

\newcommand{\bk}{\mathbf{k}}

\newcommand{\ggtext}{\hspace{0.17em}{\gg}\hspace{0.17em}}

\newcommand{\EF}{\epsilon_\text{F}}

\newcommand{\epara}{\hat{\mathbf{e}}_{\parallel}}
\newcommand{\eperp}{\hat{\mathbf{e}}_{\perp}}

\newlength\figureheight 
\newlength\figurewidth
\newcommand{\uproman}[1]{\uppercase\expandafter{\romannumeral#1}}

\begin{document}

\title{Giant DC Residual Current Generated by Subcycle Laser Pulses}


\author{Adrian Seith}
\author{Ferdinand Evers}
\author{Jan Wilhelm}
\email{jan.wilhelm@physik.uni-regensburg.de}
\affiliation{Institute  of  Theoretical  Physics and Regensburg Center for Ultrafast Nanoscopy (RUN),  University  of  Regensburg,   D-93053  Regensburg,  Germany}

\begin{abstract}
Experimental indications have been reported suggesting that laser pulses shining on materials with relativistic dispersion can produce currents that survive long after the illumination has died out.
Such residual currents (’remnants’) have applications in petahertz logical gates.
The remnants' strength strongly depends on the pulse-shape. 
We develop an analytical formula that allows to optimize the pulse-shape for remnant production; we predict remnants exceeding the values observed so far by orders of magnitude. 
In fact, remnants can be almost as strong as the peak current under irradiation.
\end{abstract}

\maketitle



The dynamics of currents induced by an ultra-short laser pulse in a metal or semiconductor is strongly influenced by the pulse's shape and amplitude. 
Improvements in pulse-shape engineering ~\cite{Krausz2014, Schubert2014, Kampfrath2013, Seifert2016, Meineke2022, Meierhofer2023} have provided access to a plethora of phenomena on subcycle time scales.
Such phenomena include the generation of high-harmonics~\cite{Chin2001, Ghimire2011, Schubert2014, Hohenleutner2015, Vampa2014, Silva2019, Schmid2021, Goulielmakis2022, Uzan-Narovlansky2022, Graml2022}, dynamical Bloch oscillations~\cite{Schubert2014, Luu2015, Borsch2020}, or ultrafast clocking of electron-electron correlations~\cite{Freudenstein2022}.
With an eye on applications it has been proposed that pulse-shape engineering and the associated control of microscopic currents should be viewed as a prerequisite of ultrafast light-wave electronics \cite{Krausz2014,Higuchi2017, Reimann2018, Heide2018, Lefebvre18, Mak2014, Moltagh2018, Motlagh2019, Galan20, Galan21, Rybka2016,Borsch2023}. 
In this context experimental observations are of relevance reporting that suitable pumping pulses can produce a DC-residual current ("remnant")~\cite{ishikawa2013, Higuchi2017, Heide2019}, which persists long after the pumping pulse has died out. 
Experiments show a strong dependence of remnants
on the pulse-shape and, in particular, on its carrier-envelope phase (CEP) \cite{Heide2021, Heide2020}, in agreement with density-matrix simulations~\cite{Wu2020}.
As a corresponding application, a petahertz logic gate~\cite{Boolakee2022} has been proposed. 
%
%

%

%
In order to use remnants as an efficient tool for light-wave driven electronics, the remnant amplitude per incoming laser pulse should be maximized, for instance to reduce energy consumption. 
Given many pulse-shape parameters 
- amplitude, carrier frequency, carrier-envelope phase, pulse-duration etc. - 
and additional material parameters, 
it is not straightforward to find an optimal setup to produce large-amplitude remnants.
Pulse-shape optimization is most conveniently achieved, if analytical expressions are known that reflect transparent parametric dependencies. 
%
While the commonly used semiclassical approximation to density-matrix dynamics yields analytic formul\ae, it fails to explain the phenomenon of remnant currents~\footnote{
In the semiclassical approximation~\cite{Xiao2010, Aversa1995, Davis1976, Chang1996}, the  time-dependent current density is given by
\begin{align*}
    \bj(t) &= q\sum_n \intbzdkpi\; \bv_n(\bk-\bA(t))\, f_n^{(0)}(\bk-\bA(t))\,,
    \\
    \hspace{1.7em}\bv_n(\bk) &\eqt  \frac{\partial \epsilon_n(\bk)}{\partial\bk}  \pt
    \bE(t)\timest\boldsymbol{\Omega}_n(\bk)\,,
    \hspace{0.3em}
    \bA(t) \eqt -q\hspace{-0.2em}{\int\limits_{-\infty}^t}\hspace{-0.2em} \bE(\tau)d\tau
    \,,
\end{align*}
where $f_n^{(0)}(\bk)$ is the initial occupation of band $n$ at crystal momentum $\bk$ in the Brillouin zone (BZ) and $\boldsymbol{\Omega}_n(\bk)$ is the Berry curvature.
For gauge consistency, the vector potential $\bA(t)$ (Coulomb gauge) and the electric field vanish for long times (see also~\cite{rauch2005}), such that we have 
\begin{align*}
\lim_{t\rightarrow\infty} \bj(t)
 = q\sum_n \intbzdkpi\,\frac{\partial \epsilon_n(\bk)}{\partial\bk}\, f_n^{(0)}(\bk)\eqqcolon \bj^{(0)}\,,
\end{align*}
where $\bj^{(0)}$ denotes the initial current before irradiation implying the absence of remnants. 
%
}; a full quantum analysis is required.

\begin{figure}[tb]
    \centering
    \includegraphics{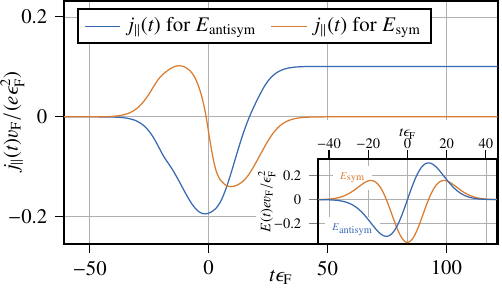}
    \caption{
    Current density~$\jpara(t)$ along the direction of the electric field in a gapless Dirac band structure. 
    The plot shows a residual current~$\dcrparallel$ that strongly depends on the incoming pulse-shape - single node $(E_\text{antisym})$ vs.~double node $(E_\text{sym})$.
    As shown, remnants can be of the same order of magnitude as the transient current.
    Inset: Electric field pulse from Eq.~\eqref{e5b} with field amplitude $E_0\eqt 0.5\,$MV/cm, pulse duration $\sigma\eqt 50$ fs leading to a carrier frequency of $\approx6.4$ THz.
    Hamiltonian parameters are chosen as typical values for a topological surface state~\cite{Wilhelm2020, Schmid2021}: Fermi-velocity $\vF\eqt 0.43\text{ nm/fs}$, Fermi energy $\EF\eqt 0.2$\,eV above the minimum of the conduction band, vanishing mass, $m_z\eqt0$, 
    implying a time scale $\EF^{-1} \eqt 3.3$ fs and a scale for the electric field amplitude  $\EF^2/(e\vF)\eqt 1.41\,\text{MV/cm}.$
    The dephasing time is $T_2\eqt1/ \gamma\eqt10$\,fs.
    }
    \label{fig1}
\end{figure}

In this work, we present an analysis that enables us to optimize parameters for finding large-amplitude remnants.
A formula is derived from third-order perturbation theory
to the time evolution mediated by the semiconductor Bloch equations (SBE)~\cite{SchmittRink1988, Aversa1995, Schaefer2002, Haug2008, Haug2009, Kira2011, Kruchinin2013, AlNaib2014, Wu2015,  Wismer2016, Li2019, Silva2019, Zhang2019, Wilhelm2020, Yue2020, Yue2022}.
%
%
In the limit of large effective damping, i.e. $\tilde{\omega}:=\omega/\gamma < 1$ with $\omega$ being the carrier frequency and $\gamma$ being the dephasing rate of the coherences, it adopts a transparent form; the remnant current density, $\dcr$, is given by 
\begin{align}
\begin{split}
    \dcr =\;& C E_0^3 \mathcal{F}[s] \big[ \epara\left(1 {+} \mathcal{O}\left(\tilde{\omega}\right)\right)+ 
    \eperp \mathcal{O}\left(\tilde{\omega}\right) \big] \\&
    +\epara\mathcal{O}(E_0^4)+\eperp\mathcal{O}(E_0^4)\,,
        \label{e2}
\end{split}
\end{align}
where $C$ is a material-dependent constant and $E_0$ is the amplitude of the linearly polarized transient electric field, 
\begin{align}
\bE(t) := E(t) \, \epara := E_0 \hspace{0.1em}s(t) \,\epara\,.\label{e1a}
\end{align}
The functional $\mathcal{F}[s]$  depends on the pulse shape  $s(t)$, $\max|s(t)|\eqt1$, while ~$\epara$ denotes a unit vector. 
We show that the value of $\mathcal{F}[s]$  strongly depends on $s(t)$;
in particular, $\dcr$ can be three orders of magnitude larger for single-cycle pulses~\cite{Meineke2022, Seifert2016} compared to  multi-cycle pulses that have previously been used for generating remnant currents in experiments~\cite{Heide2021, Heide2020, Wu2020,Boolakee2022}.
We predict that the use of single-cycle pulses~\cite{Meineke2022, Seifert2016} will result in remnants that are almost as strong as the transient currents during illumination.
We also show that expression~\eqref{e2} for remnant currents gives semi-quantitative predictions in the case of large field strength and weak damping $\omega/\gamma>1$.
\textit{Model: Time-dependent currents in SBE formalism.}
%
To describe the dynamics of non-interacting electrons, we employ the density-matrix-based SBE formalism~\cite{SchmittRink1988, Aversa1995, Schaefer2002, Haug2008, Haug2009, Kira2011, Kruchinin2013, AlNaib2014, Wu2015, Wismer2016, Li2019, Silva2019, Zhang2019, Wilhelm2020, Yue2020, Yue2022}. In the velocity gauge, the SBE are given by
\begin{align} 
\begin{split}
&\Big( \ci \frac{\partial}{\partial t} +  \ci(1-\delta_{nn'})\gamma
-\epsilon_{nn'}(\bk_t)\Big) \varrho_{nn'}(\bk, t) \\
&\phantom{\Big( \ci\frac{\partial}{\partial t} \ci ( 1}+ \ci  \delta_{nn'} \bar{\gamma}  \left( \rho_{nn}(\bk, t) - \varrho_{nn}(\bk, t_0) \right) \\
&= \bE(t) 
\sum_{\un} 
\varrho_{n\un}(\bk, t)\bfd_{\un n'}(\bk_t)
-\,\bfd_{n\un}(\bk_t)\varrho_{\un n'}(\bk, t)\,,
\end{split}\label{esbe}
\end{align} 
adopting the nomenclature of Ref.~\cite{Wilhelm2020}: $\epsilon_{n\un}(\bk) \coloneqqt \epsilon_n(\bk) \mt \epsilon_{\un} (\bk)$ is the energy difference of two bands~$n,\un$ at crystal momentum~$\bk$ in the Brillouin zone (BZ) and $\bfd_{n \un}(\bk)$ is the transition dipole matrix element. 
The crystal momentum has a time dependence stemming from minimal coupling $\bk_t := \bk + q \int_{t_0}^{t} \bE(t') \dd t' := \bk - \bA(t)$, where $\bA(t)$ can be interpreted as a vector potential.
To mimic dephasing effects of electron-electron or electron-phonon interaction, damping rates to the density-matrix have been included, where we set the diagonal damping $\bar{\gamma} = 1/T_1 = 0$
\footnote{The off-diagonal damping $\gamma\eqt 1/T_2$ (interband processes) is typically taken to be between 10 and 100\,fs,
which is of the order of the pumping frequency~\cite{Floss2018,Vampa2014,Boolakee2022,korolev2024}. 
The diagonal damping (intraband processes)  is much slower: In experiments, typical values for $T_1$ are in the order of few picoseconds for topological insulators~\cite{Kuroda2016}, which exceeds the usual pulse duration by a factor of 10 and 
is roughly 1000 times longer than the intrinsic time scale~$1/\EF\eqt 3.3$\,fs of the Hamiltonian that we have considered in Fig.~\ref{fig1}.}.

Using the time-dependent density-matrix~$\rho(t)$, we calculate the current density~$\bj(t)$ and the remnant density $\dcr$ as
\begin{align}
\bj(t) \ \coloneqq \frac{q}{V}\, \text{Tr}(\rho(t)\dot\br)\,,
\hspace{2em}   
 \dcr \ \coloneqq \underset{t\rightarrow\infty}{\lim} \bj(t)\;,
   \label{e1}
\end{align}
where $q/V$ denotes the electron charge density per unit cell volume and $\dot \br$ the velocity operator. 
%

\textit{Model: Material system.}
We focus on a discretized massive Dirac Hamiltonian 
\begin{align}
 h(\bk) := \vF(k_x\sigma_y-k_y\sigma_x) + m_z\sigma_z\, ,
 \label{e7}
\end{align}
with parameters Fermi velocity $\vF$ and mass $m_z$.
Such Dirac models have been frequently employed to describe the low-energy excitations of graphene~\cite{Castro2009}, two-dimensional semiconductors~\cite{Kormanyos2015} and topological surface states~\cite{Liu2010}. 
%
%

%

\textit{Numerical illustration.} 
Before we sketch the analytical derivation, we present a numerical simulation to illustrate remnant currents in a driven Dirac system.
To this end, the SBE~\eqref{esbe} are solved on a discretized mesh approximating the Brillouin-zone using a fourth order Runge-Kutta solver with discrete time steps.
%
%
Information about numerical convergence checks can be found in Supporting Information (SI), Sec.~\uproman{4}.
For a detailed derivation and information about the implementation of the SBE, we refer to~\cite{Wilhelm2020}.

Fig.~\ref{fig1} displays the current density parallel to the electric field for a gapless Dirac cone $(m_z=0)$, with 
$E(t)$ defined in the inset; the perpendicular current vanishes~\footnote{A gapless Dirac Hamiltonian is time-reversal symmetric and thus features no perpendicular currents, that would break time-reversal symmetry. Introducing a time-reversal breaking term, such as a gap in turn would produce nonzero currents perpendicular to the driving field.}.
As seen there, the remnant in the direction of the driving field~$\dcrparallel:= \mathbf{j}^{(\infty)}\cdott\epara$ reaches the same magnitude as the peak of $|\jpara(t)|$ and is in this sense gigantic.
It is orders of magnitude larger than the reported experimental results~\cite{ishikawa2013, Heide2019, Heide2021, Heide2020}.
The difference to the experimental situation is that we have been employing a single-cycle pulse, while experiments have been done with multi-cycle pulses. 
As we show in the following, using single-cycle pulses with a CEP leading to a suitable nodal structure is key to producing giant remnants. 
%

%
%
\textit{Analytics: Power expansion of the SBE.}
For the analysis of remnants as function of pulse-shape and material parameters, we expand the time-dependent density-matrix~$\rho(t)$, in a dimensionless parameter proportional to the electric field strength~$E_0$.

Straightforwardly, the leading term in this expansion is of third order or higher in $E_0$:
remnant amplitudes cannot be linear in the driving field. 
This is because in linear response $\mathbf{j}(t) = \int_{-\infty}^{t} \sigma(t-t')\bE(t') \dd t'$; the integral is dominated by times $t\sim t'$, because of the limited memory of the conductivity $\sigma$.
Specifically, $\mathbf{j}$ vanishes at times $t$ so large that the peak time of $\bE(t)$  has passed by more than the memory time.
Further, for the inversion- or time-reversal-symmetric models we have in mind, remnants flip their sign if $E(t)$ does, and therefore even-order terms in $E_0$ cannot arise~\footnote{Experimental measurements of the CEP dependence exhibit a sign change of remnants under electric field inversion~\cite{Heide2021, Heide2020, Boolakee2022} indicating that the symmetric situation we consider is indeed experimentally realized.}.

The formal derivation of our analytical formula for the remnant $\dcr$ has been relegated to the supplementary, Sec.~\uproman{2}.
The main steps: 
First, we expand the time evolution defined by \eqref{esbe} in the limit of weak dipole coupling $\bE(t)\cdot\mathbf{d}(\bk_t)$. Up to second order, the occupation $f_n(\bk, t)$ of the Bloch-states $\ketnk$ changes according to
\begin{align}
    &\delta f^{(2)}_{n}(\bk, t) 
    \coloneqq \sum_{\un} f_{n\un}^{(0)}(\bk)\nonumber \\
    &\times \int\limits_{t_0}^{t} \dd t_1 \int\limits_{t_0}^{t_1} \dd t_2 \Big( e^{\displaystyle t_2 (\gamma + i w_{n\un}(t_2)) - t_1 (\gamma + i w_{n\un}(t_1))} \nonumber\\
    &\times \bE(t_1)\cdot\bfd_{\un n}(\bk_{t_1}) \bE(t_2)\cdot\bfd_{n \un}(\bk_{t_2}) + \text{c.c.} \Big) \,,
    \label{e6b}
\end{align} 
with $f_{n\un}^{(0)}(\bk)\coloneqqt
f_n(\bk, t_0) \mt f_{\un}(\bk, t_0)$ the difference of the initial occupations at the starting time $t_0$ and $w_{n\un}(t) \coloneqqt \int_{t_0}^{t} \epsilon_{n\un}(\bk_{t'})\, \dd t'$ the integral of the $\bk_t$-dependent gap.
The corresponding contribution to the current reads
\begin{align}
    {{\bf j}^{(\infty)}}^{(2)}\coloneqq \lim_{t\rightarrow\infty} \sum_{n} q
    \intbzdkpi 
    \frac{\partial\epsilon_{n}({\bf k})}{\partial {\bf k}} \delta f^{(2)}_{n}(\bk, t) \,.
    \label{e7b}
\end{align}

Second, we realize that the time-integrals simplify in the Markov regime with fast damping compared to the typical rate of change of the electric field, 
i.e. $\tilde{\omega}\eqt\omega/\gamma \lltext1$ if the rate of change is characterized mainly by a single frequency $\omega$. For the series of integrals, such as displayed in \eqref{e6b}, the Markov limit implies that all times prior to $t_1$ will be replaced by $t_1$ in the integrand. The remaining integrals can be performed analytically, where each time integration produces a prefactor of order $\tilde{\omega}$, so that $\delta f^{(l)}/\delta f^{(2)} \propto \tilde{\omega}^{l-2}$ for orders $l{=}3$ or higher. Hence, terms of order three or higher in the dipole coupling are subleading in the Markov limit and will be ignored. For the second order term~\eqref{e6b} one derives 
\begin{align}
\begin{split}
        \delta f_n^{(2)}
    =\sum_{\un}f_{n\un}^{(0)}(\bk)
    \int\limits_{t_0}^{t} \dd t_1
    \dfrac{2 \big| E(t_1)  d^{\parallel}_{n \un}(\bk_{t_1}) \big|^2 }{\gamma\left(1+ \tilde{\epsilon}^2_{n\un}(\bk_{t_1}) \right) } \big(1+\mathcal{O}(\tilde{\omega})\big) \,,
    \label{eq6}
\end{split}
\end{align}
where $\tilde{\epsilon} := \epsilon/\gamma$ and $d_{n \un}^\parallel(\bk) \coloneqqt \epara \cdott \bfd_{n \un}(\bk)$. 

The driving field strength $E_0$ enters the occupation dynamics \eqref{eq6} explicitly in the denominator and implicitly in the argument ${\bf k}_t$, resulting in current contributions of second and higher order in $E_0$. 
In presence of inversion or time-reversal symmetry, the contribution of order $E_0^2$ in $\delta f^{(2)}$ is an even function of ${\bf k}$; 
one readily infers from \eqref{e7b} that the second-order remnant current vanishes, as expected, because the velocity $\partial\epsilon_n({\bf k})/\partial {\bf k}$ is odd in ${\bf k}$.

Summarizing, the surviving terms in the Markov limit lead to Eq.~\eqref{e2}, where the material constant is defined as
\begin{align}
\begin{split}
    C := \;&2 q^2 \intbzdkpi \sum_{n\un} f_{n\un}^{(0)}(\bk) \dfrac{\left( \partial_{k_\parallel} \tilde{\epsilon}_n(\bk) \right)}{1+\tilde{\epsilon}^2_{n \un}(\bk)} 
    \\
&\times\left( \partial_{k_\parallel} \big|d^\parallel_{n\un}(\bk)\big|^2 - \big|d^\parallel_{n\un}(\bk)\big|^2 \dfrac{ \partial_k \big| \tilde{\epsilon}_{n\un}(\bk) \big|^2 }{1+\tilde{\epsilon}^2_{n\un}(\bk)} \right) \,;
    \label{e3}
\end{split}
\end{align}
$\mathcal{F}[s]$ in Eq.~\eqref{e2} abbreviates the functional
\begin{align}
\mathcal{F}[s] &:= \int\limits_{t_0}^{\infty}\dd t\, ( s(t) )^2 \int\limits_{t_0}^{t} \dd t' \, s(t') \,.
 \label{e5a} 
\end{align}
\begin{figure}[ht]
    \centering
    \includegraphics{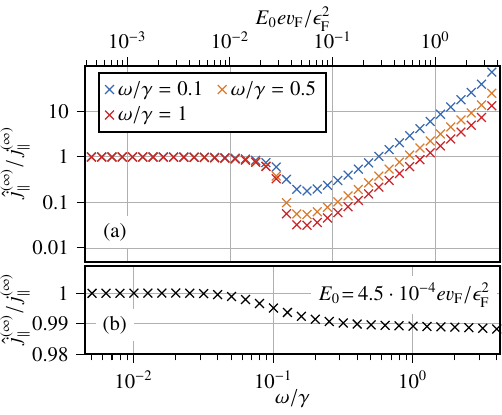}
    \caption{
    Comparison of the analytical, leading order result ~$\dcrdomparallel$ of the remnant, Eq.~\eqref{e6a}, and numerically exact solutions, $\dcrparallel$ of Eq.~\eqref{e1} for a gapless Dirac system with Fermi energy $\epsilon_\text{F}/\gamma=3$.
    (a) Ratio $\dcrdomparallel/\dcrparallel$ as a function of the field strength $E_0$ for different dephasing rates $\omega/\gamma$.
    %
    %
    The evolution of this deviation with increasing damping times $\omega/\gamma$ is plotted in (b) at $E_0=4.5\cdot10^{-4} e v_\text{F}/\epsilon^2_\text{F}$, which is deep in the cubic regime, see (a). 
    Simulation parameters are chosen as described in the caption of Fig.~\ref{fig1}.
}
    \label{fig_comp}
\end{figure}
%
%

%
Eq.~\eqref{e2} is an expression for remnants given as a double expansion in the field strength $E_0$ and the inverse effective dephasing $\tilde{\omega}$.
For small field strengths and fast dephasing relative to the electric field oscillation $\gamma>\omega$, we identify for the parallel current
\begin{align}
    \dcrdomparallel = C \mathcal{F}[s] E_0^3\label{e6a}
\end{align}
as the leading contribution to the remnants.

\textit{Numerics: Corrections to leading order} $\dcrdomparallel$.
We now discuss higher order terms by comparing the analytics from Eq.~\eqref{e6a} with numerically exact computations. 
Fig.~\ref{fig_comp} confirms the cubic field strength dependence of remnants by comparing the numerically exact data $\dcrparallel$ with the leading contribution $\dcrdomparallel$.
As shown in Fig.~\ref{fig_comp} (b), the deviation between $\dcrdomparallel$ and $\dcrparallel$ in the small field strength limit is below 2\% for all considered damping rates $0<\omega/\gamma<4$.
Fig.~\ref{fig_comp} (a) also shows, that starting at a certain field strength value, $E_0 e v_\text{F} / \EF^2 \approx 0.02$, higher order terms start to influence remnants and the approximations made in the previous sections break down.

%
%
\textit{Analytics: Optimizing pulse shapes.}
In the integrand in Eq.~\eqref{e5a}, we identify two factors:  
The first, $s(t)^2$, is non-negative. 
Further, the second factor, $\int^t_{t_0} dt' s(t') = - A(t)/(q E_0)$, represents an integral over a function, $s(t')$, that can be interpreted as a normalized vector potential. 
It can exhibit nodes, i.e. sign changes, see inset of Fig.~\ref{fig1}. 
Depending on the nodal structure, the integrand can be positive (inset, blue trace), or oscillating (inset, orange trace); 
the latter implies partial cancellations in the main integral over the time $t$ in \eqref{e5a}. 
To produce large remnants, pulse shapes with well defined polarity of the vector potential $A(t)$ are preferential in order to avoid cancellations. 
The antisymmetrically shaped electric field, blue pulse in the inset of Fig.~\ref{fig1}, corresponds to $A(t)<0$ and represents a typical example. 

In the following, we will apply Eqs.~\eqref{e2} and~\eqref{e5a} for a systematic optimization of the pulse parameters with respect to remnant production.
A general model for pulse shapes considered in~\cite{Schubert2014, Meierhofer2023, Heide2020, Boolakee2022, Graml2022, Wilhelm2020, Seifert2016,Meineke2022,Higuchi2017, Heide2019, Heide2021, Seifert2022, Lu2022} is
\begin{align}
E(t)  := E_0  \big[\cos(\omega t+\varphi) + \alpha\big]\,e^{-(t/\sigma)^2} \,,
    \label{e5}
\end{align}
with parameters CEP~$\varphi$~\footnote{In our previous work~\cite{Wilhelm2020, Graml2022, Schmid2021}, we use a sine instead of a cosine. Here, we choose the cosine in order to keep consistency with the existing literature about residual currents. To compare with our previous publications, one has to transform $\varphi\rightarrow\varphi-\pi/2$}, frequency~$\omega$ and width parameter $\sigma$~\footnote{the FWHM is given by $2\sigma\sqrt{\ln2}$}.
%
%
The factor $\alpha \coloneqqt {-}\, e^{-(\omega\sigma)^2/4}\cos(\varphi)$ is included to satisfy $\lim\limits_{t\rightarrow\infty} \int_{-\infty}^t \bE(t')dt' = 0$, which ensures gauge consistency.  
In the limit $\omega\sigma\gg 1$, Eq.~\eqref{e5} represents a  multi-cycle (mc) pulse. To model the single-cycle (sc) pulses employed in Fig.~\ref{fig1}, we evaluate Eq.~\eqref{e5} in the limit $\omega\sigma \lesssim \pi$ and then substitute $\omega\to 2/\sigma$: 
\begin{align}
E(t) = E_0
\big[2 \sin(\varphi)\,t/\sigma + \cos(\varphi)(1 - 2 (t/\sigma)^2) \big]\,e^{-(t/\sigma)^2}.
     \label{e5b}
\end{align}
The substitution ensures that the peak values of the mc- versus sc-pulses match, so that the comparison of both is meaningful.  


We evaluate the shape functional~\eqref{e5a} for the mc- and sc-pulses, Eqs.~\eqref{e5} and \eqref{e5b}, and obtain 
\begin{align}
    \mathcal{F}[s] =& \, \sigma^2 \sin(\varphi)
    \left\{
    \begin{array}{cl}
     \frac{2\sqrt{\pi}}{3\sqrt{3}} & \;\text{ for sc-pulses,} \\[1em]
     \frac{ e^{-(\omega\sigma)^2/12}}{4\sqrt{3/\pi}\,\omega\sigma} & 
     \;\text{ for mc-pulses.}
    \end{array}
    \right.  
     \label{e6}
\end{align}
As seen here, mc-pulses are exponentially suppressed as compared to sc-pulses. Further, largest remnants are produced at a CEP $\varphi=\pi/2$.
%
%

\begin{figure}[ht]
    \centering
    \includegraphics{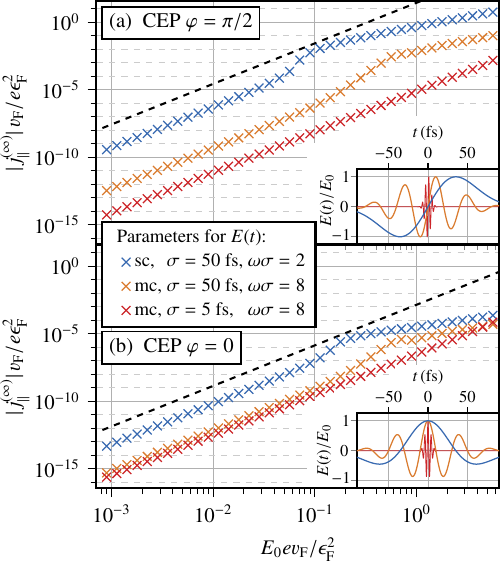}
    \caption{Remnant $\dcrparallel$ computed from SBE for varying field strengths $E_0$ for pulses with  (a) CEP $\varphi \eqt \pi/2$ and (b) $\varphi\eqt0$. 
    Dashed lines are shown as a guide for the $E_0^3$ dependence.
    Inset: Shape of electric field pulses.
    We choose a Dirac Hamiltonian with parameters as described in the caption of  Fig.~\ref{fig1}.
    Blue traces correspond to a sc-pulse  according to Eq.~\eqref{e5b}, orange and red traces to mc-pulses according to Eq.~\eqref{e5}.
    }
    \label{fig2}
\end{figure}

\textit{Numerics: Pulse shape dependencies beyond the leading order $\dcrdomparallel$}.
%
Fig.~\ref{fig2} shows numerically calculated remnants and confirms the analytical predictions made for the limit of small field strengths:
(i) As guided by the black dashed line, we observe a cubic $E_0$-dependence of remnants for all shapes in the small field strength limit. 
(ii) sc-pulses produce remnants exceeding the ones of mc-pulses by orders of magnitude. 
(iii) 
To compare the effect of different pulse lengths, we keep the number of half-cycles, i.e. $\omega\sigma$, invariant, while scaling $\sigma$.
In this procedure reducing the pulse duration by one order of magnitude reduces the remnant in Fig.~\ref{fig2} (a) by two orders of magnitude as predicted by the $\sigma^2$-dependence in Eq.~\eqref{e6}. 
(iv) For pulses with a CEP of $\varphi\eqt0$ the remnant strength is suppressed. Taking Eq.~\eqref{e6} at face value implies the absence of remnants; non-zero values observed in Fig.~\ref{fig2} (b) reflects terms of $\mathcal{O}(\omega/\gamma)$ that have been ignored when deriving Eq.~\eqref{e6}.
%
%
(v) Exploring higher field strengths, we find that the predictions made from analysing the leading terms still hold in this regime: here too, sc-pulses with a CEP of $\varphi=\pi/2$ lead to the largest remnants out of all considered shapes.
%

%
%

\textit{Materials: Remnants in topological surface states and TMDCs.}
%
For the case of a massive Dirac Hamiltonian, Eq.~\eqref{e7}, we obtain a closed expression for the material constant 
   $C \coloneqq C\left(\tilde{\epsilon}_\text{F}, \tilde {m}_z \right),$
with $\tilde{\epsilon}_\text{F} := \EF/\gamma$ and $\tilde{m}_z :=m_z/\gamma$.
The full analytical expression of $C$ is shown in the SI, Sec.~\uproman{3}.
The limiting case of zero mass $m_z=0$ and $\tilde{\epsilon}_\text{F}\gg1$ may serve to model a gapless topological surface state (TSS)~\cite{chen2009} and the material constant reads 
\begin{align}
    C^\text{TSS}  = \dfrac{q^4 \vF^2}{32 \gamma^2 \tilde{\epsilon}_\text{F}^3} + \mathcal{O}\left(\tilde{\epsilon}_\text{F}^{-4}\right) \,.
    \label{e11}
\end{align}
Note that $C^\text{TSS}$ diverges for~$\tilde{\epsilon}_\text{F}\eqt0$ signalizing a breakdown of the perturbative expansion. 
Another limit, $\tilde{m}_z\ggtext1$ and $\tilde{\epsilon}_\text{F}\eqt0$,~\footnote{For a Fermi-level in the gap, $k_\text{F}=0$. Thus, we evaluate $C$ at $\tilde{\epsilon}_\text{F}=0$.} 
models a monolayer transition metal dichalcogenide (TMDC)~\cite{Kormanyos2015};
we obtain 
\begin{align}
    C^\text{TMDC} = \dfrac{q^4 \vF^2}{35 \gamma^2 \tilde{m}_z^3} + \mathcal{O}\left(\tilde{m}_z^{-4}\right) \,.
    \label{e12}
\end{align}
To compare these materials, we consider the fraction 
\begin{align}
    \frac{\dcrdomparallel(\text{TSS})}{\dcrdomparallel(\text{TMDC})}
    = 
     \frac{C^\text{TSS}}{C^\text{TMDC}} 
  \approx    \frac{m_z^3}{\EF^3}\,,\label{e13}
\end{align}
which is independent of the pulse-shape.
Using typical parameters $\EF\eqt0.2\,$eV~\cite{chen2009}, $\gamma\eqt(10\,\text{fs})^{-1}$~\cite{Floss2018}, $m_z\eqt 2$\,eV~\cite{Kormanyos2015}, and assuming $\vF$ to be equal in the TSS and the TMDC, we estimate 
$C^\text{TSS}/C^\text{TMDC}\approx 10^3$, making a TSS (or graphene) an ideal platform to explore large remnants.
%
%

\textit{Conclusion.}
In this letter, we have performed combined analytical and numerical analyses of steady-state currents (remnants) that continue to flow in a material long after the driving laser pulse has died out. 
We predict conditions under which remnants can be of same order of magnitude as the transient current under driving. 
Moreover, remnants will be orders of magnitude larger for single-cycle pulses as compared to the multi-cycle pulses that have been used in experimental remnant studies so far. 
Our prediction relies on an analytical formula that we have derived. It explains the strong dependence of remnants on laser pulse shape and Hamiltonian parameters. 
We believe that our finding helps to boost potential applications of remnants in  ultrafast electronics.

\textit{Acknowledgements.}
We thank J.~Freudenstein, M.~Meierhofer, C.~Meinecke, J.~Schlosser for helpful discussions.
Further, we express our gratitude to P.~Hommelhoff for critical reading of our manuscript. 
Support from the German Research Foundation (DFG) through the Collaborative Research Center, Project ID 314695032 SFB 1277 (project A03) is gratefully acknowledged.
%
%
J.W.~acknowledges funding by the DFG via the Emmy Noether
Programme (Project No.~503985532). 
The authors gratefully acknowledge the scientific support and HPC resources provided by the Erlangen National High Performance Computing Center (NHR@FAU) of the Friedrich-Alexander-Universität Erlangen-Nürnberg (FAU) under the NHR project b165da10. NHR funding is provided by federal and Bavarian state authorities. NHR@FAU hardware is partially funded by the German Research Foundation (DFG) - 440719683.


\bibliographystyle{apsrev4-1}
\bibliography{Literature}

\end{document}


\title{Supplemental Material for \\[0.5em] "Giant DC Residual Current Generated by Subcycle Laser Pulses"}

\author{Adrian Seith}
\author{Ferdinand Evers}
\author{Jan Wilhelm}
\email{jan.wilhelm@physik.uni-regensburg.de}
\affiliation{Institute  of  Theoretical  Physics  and Regensburg Center for Ultrafast Nanoscopy (RUN),  University  of  Regensburg,   D-93053  Regensburg,  Germany}

\begin{abstract}
The supplemental material with additional information to 'Giant DC Residual Current Generated by Subcycle Laser Pulses' contains an explanation on the natural units used in the main part in Sec.~\ref{app:A}. Sec.~\ref{app:B} lays out the detailled derivation for the analytical formula describing remnant currents. In Sec.~\ref{app:D}, the full analytical formula for the material-dependent constant will be given for a massive Dirac Hamiltonian. Lastly, in Sec.~\ref{app:F}, the computational details for all calculations used in this publication are given.
\end{abstract}

\maketitle

\section{Natural Units of Remnant Currents}\label{app:A}
In all calculations, we have set $\hbar = 1$. As a natural unit for driven currents in a Dirac system, we relate the current density $j(t)$ to the Fermi velocity $v_\text{F}$ and the Fermi energy $\epsilon_\text{F}$ of the problem. The intraband current, see Eq.~\eqref{ec2}, is given by a Brillouin zone integral that can be compared with the typical length scale of the problem, which is the Fermi-vector $k_\text{F}$ times the derivative of the bandstructure, that is $v_\text{F}$ for a massless Dirac system and the charge of the electron. Therefore, a natural way to express the current response is $j^{(\infty)}v_\text{F}/(e\epsilon_\text{F}^2)$. The electric field as the time derivative of the vector potential has an additional frequency (energy) dependence, that can be expressed by $\epsilon_\text{F}$ and thus is expressed as $e E(t)v_\text{F} / \epsilon_\text{F}^2$. 

\section{Remnant Currents from a perturbative expansion of the Semiconductor Bloch Equations}\label{app:B}
The Semiconductor Bloch equations (SBEs) in the 'Coulomb' or 'velocity' gauge with a phenomenological, off-diagonal damping $\gamma$, read~\cite{Wilhelm2020}
\begin{align} 
\begin{split}
\Big( \ci \frac{\partial}{\partial t} +  \ci(1-\delta_{nn'})\gamma
-\epsilon_{nn'}(\bk_t)\Big)
\varrho_{nn'}(\bk,t) = \bE(t) 
\sum_{\un} 
\varrho_{n\un}(\bk,t)\bfd_{\un n'}(\bk_t)
-\,\bfd_{n\un}(\bk_t)\varrho_{\un n'}(\bk,t)\,.
\end{split}\label{eq:sbe}
\end{align} 
where the initial (unperturbed) Bloch Hamiltonian $h(\bk)$ defines the eigenenergies and lattice-periodic part of the Bloch eigenstates of the system via
\begin{align}
h(\bk) \,|n\bk\brangle = 
\epsilon_{n}(\bk) \,|n\bk\brangle\,,
\end{align}
and 
\begin{equation}
\epsilon_{n\un}(\bk_t) \coloneqq \epsilon_n(\bk_t) - \epsilon_{\un}(\bk_t)
\end{equation}
is the difference of the bandstructure of bands $n$ and $\un$. 
They describe the time dynamics for the density-matrix in the Bloch basis
\begin{equation} 
    \varrho_{n\un}(\bk, t) = \blangle n \bk | \varrho(t) | \un \bk \brangle
\end{equation}
as a response to an homogeneous external driving field $\bE(t)$ with ~$\underset{t\rightarrow-t_0}{\lim}\bE(t) = 0$, that acts as a time-dependent perturbation. The driving field enters the SBEs in this gauge via minimal coupling
\begin{align}
    \bk \rightarrow \bk_t := \bk + q \int_{-t_0}^{t} \dd t'\, \bE(t') := \bk - \bA(t) \,,
\end{align}
as well as via a coupling of diagonal and off-diagonal elements of the density-matrix by the dipole matrix elements defined by
\begin{align}
\bfd_{n\un}(\bk_t) &=  \cq \bracom\ci\partial_{\bk_t}\ketcomprime\,\label{e32}.
\end{align}
For a detailed derivation of the SBE formalism, we refer to Ref.~\cite{Wilhelm2020}.

By writing the density-matrix as a vector, both sides of the SBEs can be expressed as matrix-vector products. For a two-band system, with
\begin{align}
    \boldsymbol{\varrho}(\bk, t) &:= \begin{pmatrix} \varrho_{cc}(\bk ,t) \\ \varrho_{vv}(\bk ,t) \\ \varrho_{cv}(\bk ,t) \\ \varrho_{vc}(\bk ,t) \end{pmatrix} = \begin{pmatrix} f_c(\bk ,t) \\ f_v(\bk ,t) \\ \varrho_{cv}(\bk ,t) \\ \varrho_{vc}(\bk ,t) \end{pmatrix} ,
    \label{eq:B7} \hspace{5em}
    \mathds{H}(\bk_t) := \begin{pmatrix}
        0 & 0 & 0 & 0 \\
        0 & 0 & 0 & 0 \\
        0 & 0 & \epsilon_{cv}(\bk_t) & 0 \\
        0 & 0 & 0 &  \epsilon_{vc}(\bk_t) 
    \end{pmatrix}, \\
    \boldsymbol{\gamma} &:= \begin{pmatrix}
        0 & 0 & 0 & 0 \\
        0 & 0 & 0 & 0 \\
        0 & 0 & \gamma & 0 \\
        0 & 0 & 0 & \gamma 
    \end{pmatrix}, \hspace{2em}
    \mathds{V}(\bk_t, t) := \bE(t) \cdot \begin{pmatrix}
        0 & 0 & \bfd_{vc}(\bk_t)  & -\bfd_{cv}(\bk_t)  \\
        0 & 0 & -\bfd_{vc}(\bk_t)  & \bfd_{cv}(\bk_t)  \\
        \bfd_{cv}(\bk_t)  & -\bfd_{cv}(\bk_t)  & \bfd_{vv}(\bk_t)  -\bfd_{cc}(\bk_t)  & 0 \\
        - \bfd_{vc}(\bk_t)  & \bfd_{vc}(\bk_t)  & 0 & \bfd_{cc}(\bk_t) - \bfd_{vv}(\bk_t) 
    \end{pmatrix}.
\end{align}
the SBEs in matrix form are:
\begin{align}
    \ci \frac{\partial}{\partial t} \boldsymbol{\varrho}(\bk, t) = \Big( \mathds{H}(\bk_t) - \ci \boldsymbol{\gamma} + \mathds{V}(\bk_t, t) \Big) \boldsymbol{\varrho}(\bk, t)
    \label{eq:b3}
\end{align}
To perform a perturbative expansion in $\bE(t)$, we first make the ansatz 
\begin{align}
    \boldsymbol{\varrho}(\bk, t) = \begin{pmatrix} \varrho_{cc}(\bk ,t) \\ \varrho_{vv}(\bk ,t) \\ \tilde{\varrho}_{cv}(\bk ,t) e^{-\gamma t}\\ \tilde{\varrho}_{vc}(\bk ,t) e^{-\gamma t} \end{pmatrix}, 
    \label{eq:B12}
\end{align}
which, inserted into Eq.~\eqref{eq:b3} leads to the following equation of motion:
\begin{align}
    \ci \frac{\partial}{\partial t} \tilde{\boldsymbol{\varrho}}(\bk, t) = \Big( \mathds{H}(\bk_t) + \bE(t)\tilde{\mathds{V}}(\bk, t) \Big) \tilde{\boldsymbol{\varrho}}(\bk, t)\,,
\end{align}
whose expressions for two-band models are given by
\begin{align}
    \tilde{\boldsymbol{\varrho}}(\bk, t) &:= \begin{pmatrix} \varrho_{cc}(\bk ,t) \\ \varrho_{vv}(\bk ,t) \\ \tilde{\varrho}_{cv}(\bk ,t) \\ \tilde{\varrho}_{vc}(\bk ,t) \end{pmatrix}, \hspace{1em}
    \tilde{\mathds{V}}(\bk, t) := \bE(t) \cdot \begin{pmatrix}
        0 & 0 & \bfd_{vc}(\bk_t) e^{-\gamma t} & -\bfd_{cv}(\bk_t) e^{-\gamma t} \\
        0 & 0 & -\bfd_{vc}(\bk_t) e^{-\gamma t} & \bfd_{cv}(\bk_t) e^{-\gamma t} \\
        \bfd_{cv}(\bk_t) e^{\gamma t} & -\bfd_{cv}(\bk_t) e^{\gamma t} & \bfd_{vv}(\bk_t)  -\bfd_{cc}(\bk_t)  & 0 \\
        - \bfd_{vc}(\bk_t) e^{\gamma t} & \bfd_{vc}(\bk_t) e^{\gamma t} & 0 & \bfd_{cc}(\bk_t) - \bfd_{vv}(\bk_t) 
    \end{pmatrix}.
\end{align}
In the Dirac-picture, the EOM has the form
\begin{align}
    \ci \frac{\partial}{\partial t} \tilde{\boldsymbol{\varrho}}_I(\bk, t) = \tilde{\mathds{V}}_I(\bk, t)\tilde{\boldsymbol{\varrho}}_I(\bk, t) \,,
\label{eq:rhodirac}
\end{align}
with
\begin{align}
    \tilde{\boldsymbol{\varrho}}_I(\bk, t) &:= e^{\ci \int_{t_0}^t \mathds{H}(\bk_{t'}) \dd t'} \tilde{\boldsymbol{\varrho}}(\bk, t), \label{eq:B17} \hspace{5em}
    \tilde{\mathds{V}}_I(\bk, t) := e^{\ci \int_{t_0}^t \mathds{H}(\bk_{t'}) \dd t'} \tilde{\mathds{V}}(\bk, t) e^{ - \ci \int_{t_0}^t \mathds{H}(\bk_{t'}) \dd t'} . 
\end{align}
Here, we can neglect the time-ordering operator, because $\left[\mathds{H}(\bk_t), \mathds{H}(\bk_{t'})\right] = 0$.
Defining $w_{ij}(t) := \int_{t_0}^{t} \epsilon_{ij}(\bk_{t'}) \dd t'$, the interaction matrix can be written as
\begin{align}
    \tilde{\mathds{V}}_I(\bk, t) = \bE(t) \cdot \begin{pmatrix}
        0 & 0 & \bfd_{vc}(\bk_t) e^{-\gamma t - \ci w_{vc}(t)} & -\bfd_{cv}(\bk_t) e^{-\gamma t - \ci w_{cv}(t)} \\
        0 & 0 & -\bfd_{vc}(\bk_t) e^{-\gamma t - \ci w_{vc}(t)} & \bfd_{cv}(\bk_t) e^{-\gamma t - \ci w_{cv}(t)} \\
        \bfd_{cv}(\bk_t) e^{\gamma t + \ci w_{cv}(t)} & -\bfd_{cv}(\bk_t) e^{\gamma t + \ci w_{cv}(t)} & \bfd_{vv}(\bk_t)  -\bfd_{cc}(\bk_t)  & 0 \\
        - \bfd_{vc}(\bk_t) e^{\gamma t + \ci w_{vc}(t)} & \bfd_{vc}(\bk_t) e^{\gamma t + \ci w_{vc}(t)} & 0 & \bfd_{cc}(\bk_t) - \bfd_{vv}(\bk_t) 
    \end{pmatrix}.
\end{align}
The initial condition of the system is given by $\tilde{\varrho}_{I, ij}^{(0)}(\bk) = \varrho_{ij}^{(0)}(\bk)\delta_{ij}$.
Time integration of Eq.~\eqref{eq:rhodirac} and iterative insertion of $\tilde{\boldsymbol{\varrho}}_I(\bk, t)$ leads to a series expansion in $\mathbf{E}\cdot\mathbf{d}$.
\begin{align}
\begin{split}
    \tilde{\boldsymbol{\varrho}}_I(\bk, t) &= \tilde{\boldsymbol{\varrho}}_I(\bk, t_0) + \dfrac{1}{i} \int_{t_0}^{t} \dd t' \tilde{\mathds{V}}_I(\bk,  t') \tilde{\boldsymbol{\varrho}}_I(\bk, t_0) + \dfrac{1}{i^2} \int_{t_0}^{t} \dd t' \int_{t_0}^{t'} \dd t''\tilde{\mathds{V}}_I(\bk,  t') \tilde{\mathds{V}}_I(\bk, t'') \tilde{\boldsymbol{\varrho}}_I(\bk, t_0) + ...
    \label{eq:expansion}
\end{split}
\end{align}
Embarking on Eq.~\eqref{eq:B12}, Eq.~\eqref{eq:B17} and Eq.~\eqref{eq:expansion}, we obtain the following equation for the diagonal entries of the density-matrix in the original basis \eqref{eq:B7} up to second order in the perturbation $\tilde{\mathds{V}}_I$:
\begin{align}\begin{split}
    \varrho_{nn}(\bk, t) 
    = &\varrho_{nn}(\bk, t_0) + \sum_{\un} \left(\varrho_{\un\un}(\bk, t_0) - \varrho_{nn}(\bk, t_0) \right) \\
    &\times \Bigg[ \int\limits_{t_0}^{t} \dd t_1 \int\limits_{t_0}^{t_1} \dd t_2 \left( e^{ \left( t_2 - t_1 \right) \gamma + i \left( w_{n\un}(t_2) - w_{n\un}(t_1) \right)} \bE(t_1)\cdot\bfd_{\un n}(\bk_{t_1}) \bE(t_2)\cdot\bfd_{n \un}(\bk_{t_2}) + \text{c.c.}\right) \\
    & + \mathcal{O}\left(\left(\bE \cdot \bfd\right)^3\right) \Bigg]
    \label{eq:15}
\end{split}\end{align}
For sufficiently large damping $\gamma$, only times where $t_2$ is close to $t_1$ are contributing to the $t_2$ integral. Expanding $w_{ij}(t_2)-w_{ij}(t_1)$ in the exponent for small $\Delta t := t_2 - t_1$ enables us to evaluate the $t_2$-integral in the leading order in $\Delta t$.
To do so, we expand the exponent around $\Delta t$:
\begin{align}\begin{split}
    &\int\limits_{t_0}^{t} \dd t_1 \int\limits_{t_0}^{t_1} \dd t_2 e^{ \left( t_2 - t_1 \right) \gamma + i \left( w(t_2) - w(t_1) \right)} f(t_1, t_2) 
    = \int\limits_{t_0}^{t} \dd t_1 \int\limits_{t_0-t_1}^{0} \dd \Delta t e^{ \Delta t \gamma + i \left( w(\Delta t + t_1) - w(t_1) \right) + \ln\left(f(t_1, \Delta t + t_1) \right) } \\
    &= \int\limits_{t_0}^{t} \dd t_1 \int\limits_{t_0-t_1}^{0} \dd \Delta t e^{ \Delta t \left( \gamma + i \dot{w}(t_1) \right) + \frac{1}{2} i \ddot{w}(t_1) \Delta t^2 + \ln\left(f(t_1, t_1) +  \partial_{\Delta t} \ln\left( f(t_1, t_1 + \Delta t )_{|\Delta t = 0} \right) \Delta t + \frac{1}{2} \partial^2_{\Delta t} \ln\left( f(t_1, t_1 + \Delta t ) \right)^2_{|\Delta t = 0} \Delta t^2 + \mathcal{O}\left(\Delta t^3\right) \right)}\,.
\end{split}\end{align}
Introducing 
\begin{align}
\gamma^{(1)}_\text{eff}(t_1) &:= \gamma + i \dot{w}(t_1) + \partial_{\Delta t} \ln\left( f(t_1, t_1 + \Delta t )_{|\Delta t = 0} \right) \,, \\
\gamma^{(2)}_\text{eff}(t_1) &:= \dfrac{1}{2} \left( i \ddot{w}(t_1) + \partial^2_{\Delta t} \ln\left( f(t_1, t_1 + \Delta t ) \right)^2_{|\Delta t = 0} \right)\,,
\end{align}
we can write the integral as
\begin{align}\begin{split}
    &\int\limits_{t_0}^{t} \dd t_1 \int\limits_{t_0-t_1}^{0} \dd \Delta t e^{\gamma^{(1)}_\text{eff}(t_1) \Delta t + \gamma^{(2)}_\text{eff} \Delta t^2 + \mathcal{O}(\Delta t^3)} f(t_1, t_1) \\
    &= \int\limits_{t_0}^{t} \dd t_1 \dfrac{1 - e^{ \left( t_0 - t_1 \right) \gamma^{(1)}_\text{eff}(t_1)}}{\gamma^{(1)}_\text{eff}(t_1)} f(t_1, t_1) \left( 1 + \mathcal{O}\left(|\gamma^{(2)}_\text{eff}|/|\gamma^{(1)}_\text{eff}|^2\right)\right)  \\
    &=\int\limits_{t_0}^{t} \dd t_1 \dfrac{f(t_1, t_1)}{\gamma^{(1)}_\text{eff}} \left( 1 + \mathcal{O}\left(|\gamma^{(2)}_\text{eff}|/|\gamma^{(1)}_\text{eff}|^2\right)\right) \,.\label{e20}
\end{split}\end{align}
For a large $\gamma^{(1)}_\text{eff}$, the factor $e^{(t_0-t_1)\gamma^{(1)}_\text{eff}(t_1)}$ has a sharp peak around $t_1=t_0$ and is negligible for $(t_1-t_0)\gamma^{(1)}_\text{eff}(t_1)\gg1$. $f(t_1, t_1)$ is a function of the electric field pulse, that starts at $t_0$. 
If we assume a slowly increasing pulse, we have $f\approx0$ for those times, where the exponential function is not negligible. Thus, if the peaks of $e^{(t_0-t_1)\gamma^{(1)}_\text{eff}(t_1)}$ and $f(t_1, t_1)$ are well separated, the integral containing the exponential in Eq.~\eqref{e20} is negligible compared to the integral of $f(t_1,t_1)/\gamma^{(1)}_\text{eff}$, justifying the last step in Eq.~\eqref{e20}. 
This is the case for all pulses that we consider in the main part of the manuscript.
Lastly, we can expand $1/\gamma^{(1)}_\text{eff}$ in the limit of large $\gamma$, utilizing that $\partial_t f(t,t) \propto \omega $, which is the typical frequency of the driving field. 
Keeping only leading terms in $\omega/\gamma$, as terms of $ \mathcal{O}(|\gamma^{(2)}_\text{eff}|/|\gamma^{(1)}_\text{eff}|^2)$ are also at least of $\mathcal{O}\left(\omega/\gamma\right)$,we can simplify the remaining integral by
\begin{align}\begin{split}
    \int\limits_{t_0}^{t} \dd t_1 \dfrac{f(t_1, t_1)}{\gamma^{(1)}_\text{eff}(t_1)} \left( 1 + \mathcal{O}\left(|\gamma^{(2)}_\text{eff}|/|\gamma^{(1)}_\text{eff}|^2\right)\right) = \int\limits_{t_0}^{t} \dd t_1 \dfrac{f(t_1, t_1)}{\gamma + \ci \dot{w}(t_1)} \left( 1 + \mathcal{O}\left(\omega/\gamma\right)\right)\,.
\end{split}
\end{align}
In the case of zero offdiagonal damping $(\gamma=0)$, we cannot perform the simplifications done in the last step and have to use $\gamma^{(1)}_\text{eff}(t_1)$ to approximate the density-matrix elements.
For the remainder of the paper however, we assume a nonzero damping.
Summarizing, we can approximate the diagonal elements of the density-matrix for large damping $\gamma$ up to second order in the dipole-coupling $\bE\cdot\bfd$ as
\begin{align}\begin{split}
    \varrho_{nn}(\bk, t) 
    = &\varrho_{nn}(\bk, t_0) + \sum_{\un} \left(\varrho_{\un\un}(\bk, t_0) - \varrho_{nn}(\bk, t_0) \right) \Bigg[ \int\limits_{t_0}^{t} 
    \dfrac{2 \gamma\ \bE(t_1)\cdot\bfd_{\un n}(\bk_{t_1}) \bE(t_1)\cdot\bfd_{n \un}(\bk_{t_1})}{\gamma^2 + \epsilon_{n\un}^2(\bk_{t_1}) } \Big(1+\mathcal{O}(\omega/\gamma)\Big) \dd t_1 \\
    & + \mathcal{O}\left(\left(\bE \cdot \bfd\right)^3\right)  \Bigg] \,.
    \label{eq16}
\end{split}\end{align}
For a linearly polarized field $\bE(t) = E(t) \epara$, we can expand Eq.~\eqref{eq16} around small field strengths to obtain the occupations up to cubic order in $E_0$:
\begin{align}\begin{split}
    \varrho_{nn}(\bk, t) 
    = &\varrho_{nn}(\bk, t_0) + \sum_{\un} \left(\varrho_{\un\un}(\bk, t_0) - \varrho_{nn}(\bk, t_0) \right) \Bigg[ \int\limits_{t_0}^{t} 
    \dfrac{2 \gamma}{\gamma^2 + \epsilon^2_{n\un}(\bk)}
    \Bigg(2 \big|E(t_1)\big|^2 \big|d^\parallel_{n\un}(\bk)\big|^2 \\
    &+ \big|E(t_1)\big|^2 A(t_1) \left( \partial_{k_\parallel} \big|d^\parallel_{n\un}(\bk)\big|^2 - \big|d^\parallel_{n\un}(\bk)\big|^2 \dfrac{ \partial_k \big| \epsilon_{n\un}(\bk) \big|^2 }{\gamma^2+\epsilon^2_{n\un}(\bk)} \right)\Bigg)
     \Big(1+\mathcal{O}(\omega/\gamma)\Big) \dd t_1
+ \mathcal{O}\left(E_0^4\right) \Bigg] \,.
    \label{eb20}
\end{split}\end{align}
This power-law expansion however only holds true, if the perturbation $\bE(t)\cdot \bd_{n\un}$ can be assumed to be small. This leads to the constraint on the Hamiltonian, that the dipole moments need to be free of any divergence.
Similar expansions of SBEs have already been employed earlier to calculate higher-order responses to continuous driving~\cite{mele2000coherent, aversa1995nonlinear, cheng2014third, mikhailov2016quantum}.


As described in Ref.~\cite{Wilhelm2020}, the time dependent current-density can be calculated from the density-matrix formalsim via 
\begin{align}
 \bj(t)  = \cq \sum_{n\un} \ \intbzdkpi \branksubt  \partial_{\bk_t} h^\text{in}(\bk_t)\ketnkprimesubt\ \varrho_{n\un}(\bk,t)\,.
\end{align} 
We assume a system with zero damping of band occupations but non-vanishing dephasing, i.e. damping of off-diagonal density-matrix elements.
In this case, only intraband currents can contribute to remnants, therefore in our analysis, we focus on
\begin{align}\begin{split}
     \bj^{(\infty)} \coloneqqt \underset{t\rightarrow\infty}{\lim} \bj(t) 
     = \cq
  \sum_{n} \intbzdkpi\ \partial_{\bk}\epsilon_n(\bk)\
  \varrho_{nn}(\bk,t\rightarrow\infty)\,.\label{ec2}
\end{split}\end{align}
Embarking on Eq.~\eqref{eb20}, we see that for a linearly polarized pulse with $\bA(t\rightarrow\infty) = 0$, all contributions to the remnants in pulse direction up to cubic order are
\begin{align}\begin{split}
    \bj^{(\infty)} = &q \sum_{n \un} \intbzdkpi\ \partial_{\bk}\epsilon_n(\bk)\left(\varrho_{\un\un}(\bk, t_0) - \varrho_{nn}(\bk, t_0) \right)\Bigg[\int\limits_{t_0}^{t} 
    \dfrac{2 \gamma}{\gamma^2 + \epsilon^2_{n\un}(\bk)}
    \Bigg(2 \big|E(t_1)\big|^2 \big|d^\parallel_{n\un}(\bk)\big|^2 \\
    &+ \big|E(t_1)\big|^2 A(t_1)\left( \partial_{k_\parallel} \big|d^\parallel_{n\un}(\bk)\big|^2 - \big|d^\parallel_{n\un}(\bk)\big|^2 \dfrac{ \partial_k \big| \epsilon_{n\un}(\bk) \big|^2 }{\gamma^2+\epsilon^2_{n\un}(\bk)} \right) \Bigg)
     \dd t_1
    \Big(1+\mathcal{O}(\omega/\gamma)\Big) + \mathcal{O}\left(E_0^4\right) \Bigg]\,.
    \label{eq:C3}
\end{split}\end{align}
If we assume the band structure of the system to be symmetric wrt.~sign change of $\bk$, and the dipole elements are either symmetric or antisymmetric wrt sign change in $\bk$, one directly sees, that all cubic contributions to the remnant currents vanish.
This is the case if our Hamiltonian is either time-reversal or inversion symmetric or both. In both cases, we have $\epsilon_n(\bk) = \epsilon_n(-\bk)$. 
For a time reversal symmetric Hamiltonian, we can fix the global phase such that $|n, -\bk\brangle =  | n \bk \brangle^*$ and accordingly for a inversion symmetric Hamiltonian, $|n, -\bk\brangle = |n, \bk\brangle$ up to a global phase. 
Thus $\bfd_{n\un}(-\bk) = - \bfd_{n\un}(\bk)$ and the second order terms in Eq.~\eqref{eq:C3} vanish.
A similar argument can be made for remnant contributions perpendicular to the driving field: $\partial_{k_\perp} \epsilon_n(\bk)$ is antisymmetric with respect to sign change of $k_\perp$, whereas the rest of the integrand in Eq.~\eqref{eq:C3} is symmetric wrt. sign change of $k_\perp$.
This has the consequence, that all contributions to remnants perpendicular to the driving have to be of $\mathcal{O}(\omega/\gamma)$ as compared to parallel remnants or have to be quartic in $E_0$.
Summarizing, Eq.~\eqref{eq:C3} can be written as (cf. Eq. (1) in the main text)
\begin{align}
\begin{split}
        \dcr &= E_0^3 C \mathcal{F}[s] \big( \epara + (\epara\mathcal{O}(\omega/\gamma) + \eperp\mathcal{O}(\omega/\gamma) \big) +\epara\mathcal{O}(E_0^4)+\eperp\mathcal{O}(E_0^4)\,,
\label{eq:dcr}
\end{split}
\end{align}
with
\begin{align}
\mathcal{F}[s] &:= \int\limits_{-\infty}^{\infty}\dd t\, ( s(t) )^2 \int\limits_{-\infty}^{t} \dd t' \, s(t') \,,
 \label{e5a}
\end{align}
and
\begin{align}
\begin{split}
    C \coloneqq \, 2q^2 \intbzdkpi &\sum_{n } \sum_{\un\neq n}   \dfrac{\gamma}{\gamma^2 +   \epsilon^2_{n\un}(\bk)} \left(\varrho_{\un\un}(\bk, t_0) - \varrho_{nn}(\bk, t_0) \right)
 \left( \partial_{k_\parallel}\epsilon_{n}(\bk) \right) \\
&\times\left( \partial_{k_\parallel} \big|d^\parallel_{n\un}(\bk)\big|^2 - \big|d^\parallel_{n\un}(\bk)\big|^2 \dfrac{ \partial_k \big| \epsilon_{n\un}(\bk) \big|^2 }{\gamma^2+\epsilon^2_{n\un}(\bk)} \right) \,.
    \label{ec8}
\end{split}
\end{align}
%
%
For certain Hamiltonians, that have divergent dipole matrix elements, the series expansion breaks down. This is, for example, the case for a massless Dirac cone, whose dipole elements show divergences at the $\Gamma$ point. In Eq.~\eqref{ec8}, this manifests as a divergent BZ integral. For the purposes of calculating DC remnants, we can circumvent this problem by introducing a Fermi level, that changes the initial occupations. If the conduction band of the Dirac cone is occupied, the factor $( \varrho^{(0)}_{nn}(\bk) - \varrho^{(0)}_{\un \un}(\bk) )$ is zero for $|\bk| \leq |\bk_F|$, meaning that we can exclude the Fermi circle of the BZ integral and as such also the divergent terms. 

\section{Material dependent constant for Dirac Hamiltonians}\label{app:D}
In this section, we evaluate the material dependent constant $C$ from Eq.~\eqref{eq:dcr} for a massive Dirac cone Hamiltonian 
\begin{align}
 h(\bk) := \vF(k_x\sigma_y-k_y\sigma_x) + m_z\sigma_z\,,
 \label{e7}
\end{align}
At zero temperature, the initial occupation $\varrho_{cc}{(\bk, t_0)}$ is either 1, if $|\bk|\leq k_\text{F}$ or 0 if $|\bk|>k_\text{F}$ and $\varrho_{vv}(\bk, t_0) = 1$. As all terms in Eq.~\eqref{ec8} are proportional to  $\left(\varrho_{vv}(\bk, t_0) - \varrho_{cc}(\bk, t_0) \right)$, we can transform the BZ-integral into polar coordinates and set the lower limit of the radial integral to $k_\text{F}$. 
Substituting $\tilde{\epsilon}_\text{F} = k_\text{F} \vF / \gamma $ and $\tilde{m}_z = m_z / \gamma $, we obtain a closed expression for $C$:
\begin{align}
\begin{split}
    C(\tilde{\epsilon}_\text{F},\tilde{m}_z) =& \dfrac{q^4 v_\text{F}^2}{240 \gamma^2 \left( \EFg^2 + \mzg^2 \right)^3 \left( 1 + 4 \EFg^2 + 4 \mzg^2 \right) \pi}
    \Bigg( -160 \EFg^2 \mzg^2 \sqrt{\EFg^2+\mzg^2}-2880 \mzg^{10} \sqrt{\EFg^2+\mzg^2}\\
    &-8640 \EFg^2 \mzg^8 \sqrt{\EFg^2+\mzg^2}+480 \EFg^2 \mzg^6 \sqrt{\EFg^2+\mzg^2}-76 \mzg^6 \sqrt{\EFg^2+\mzg^2}-356 \EFg^2 \mzg^4 \sqrt{\EFg^2+\mzg^2} \\
    &-64 \mzg^4 \sqrt{\EFg^2+\mzg^2}+15 \pi  \left(48 \mzg^4-8 \mzg^2+3\right) \left(\EFg^2+\mzg^2\right)^3 \left(4 \EFg^2+4 \mzg^2+1\right)\\ 
    &-30 \left(48 \mzg^4-8 \mzg^2+3\right) \left(\EFg^2+\mzg^2\right)^3 \left(4 \EFg^2+4 \mzg^2+1\right) \tan ^{-1}\left(2 \sqrt{\EFg^2+\mzg^2}\right) \\
    &+480 \EFg^6 \mzg^2 \sqrt{\EFg^2+\mzg^2}-180 \EFg^6 \sqrt{\EFg^2+\mzg^2}-2880 \EFg^6 \mzg^4 \sqrt{\EFg^2+\mzg^2}-460 \EFg^4 \mzg^2 \sqrt{\EFg^2+\mzg^2} \\
    &-60 \EFg^4 \sqrt{\EFg^2+\mzg^2}-8640 \EFg^4 \mzg^6 \sqrt{\EFg^2+\mzg^2}+960 \EFg^4 \mzg^4 \sqrt{\EFg^2+\mzg^2} \Bigg) \,.
\end{split}
\end{align}
For a gapless Dirac cone, we have
\begin{align}
    C(\tilde{\epsilon}_\text{F},\tilde{m}_z = 0) =& \dfrac{q^4 v_F^2}{16 \gamma^2 \EFg \left( 1 + 4 \EFg^2 \right) \pi} \left( 12 \pi  \EFg^3-6 \left(4 \EFg^3+\EFg\right) \tan ^{-1}(2 \EFg)-12 \EFg^2+3 \pi  \EFg-4 \right) \,.
\end{align}

\section{Computational Details}\label{app:F}
\begin{figure}[h!]
    \centering
    \includegraphics{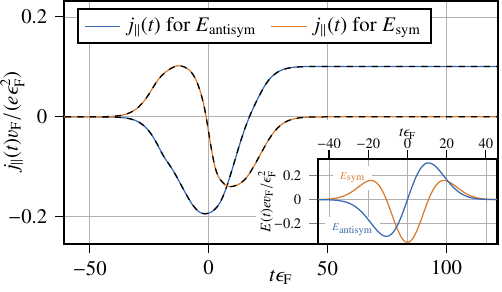}
    \caption{Convergence check for main-text Fig.~1: Coloured lines are the data plotted in the main part with $800\times144$ points, a BZ size of $1.89\text{\AA}^{-1} \times 0.95 \text{\AA}^{-1}$ and an integration time step of $0.01 \text{fs}$, dashed lines are overconverged data with $1080\times288$ points, a BZ size of $2.27\text{\AA}^{-1} \times 1.13\text{\AA}^{-1}$ and an integration time step of $0.008\,\text{fs}$.}
    \label{fig01_convergence}
\end{figure}

\begin{figure}[h!]
\begin{minipage}{.495\textwidth}
    \centering
    \includegraphics{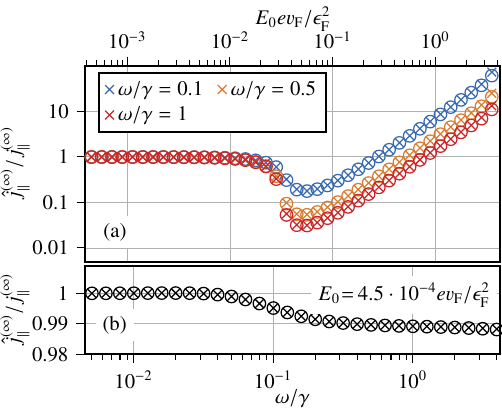}
\end{minipage}%
\begin{minipage}{.495\textwidth}
    \centering
    \includegraphics{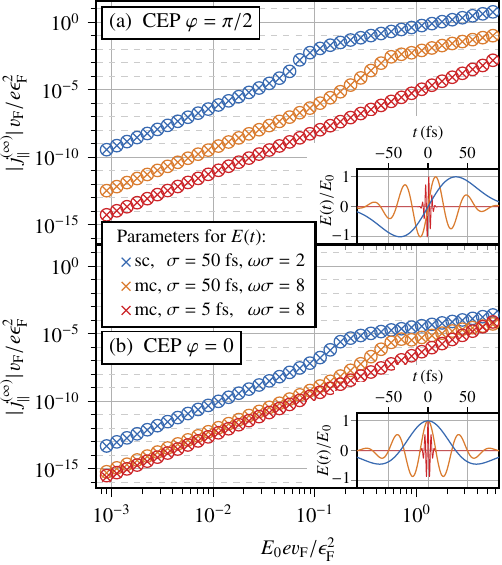}
\end{minipage}
    \caption{Convergence check for Figure 2 and 3 from the main manuscript: Crosses are the data plotted in the main part with $800\times360$ points and a BZ size of $1.89\text{\AA}^{-1} \times 0.95 \text{\AA}^{-1}$ for $E_0\leq 1.0\,\text{MV/cm}$ and $2400\times504$ points and a BZ size of $11.34\text{\AA}^{-1} \times 2.65 \text{\AA}^{-1}$ for $E_0 > 1.0\,\text{MV/cm}$.
    The integration time step for all pulses with width $50\,\text{fs}$ and CEP $\varphi=0$ was $0.01\,\text{fs}$ and $0.005\,\text{fs}$ for all other pulses. 
    Circles are overconverged with $1200\times504$ points and a BZ size of $2.27\text{\AA}^{-1} \times 1.13\text{\AA}^{-1}$ for $E_0\leq 1.0\,\text{MV/cm}$ and $3000\times648$ points and a BZ size of $13.23\text{\AA}^{-1} \times 3.02\text{\AA}^{-1}$ for $E_0 > 1.0\,\text{MV/cm}$.
    The integration time step for all pulses with width $50\,\text{fs}$ and CEP $\varphi=0$ was $0.008\,\text{fs}$ and $0.004\,\text{fs}$ for all other pulses. }\label{fig023_convergence}
\end{figure}

For the Brillouin zone sampling, we employ Monkhorst-Pack meshes~\cite{Monkhorst1976}. 
We carefully checked convergence of the Monkhorst-Pack mesh, reaching converged results for meshes with a density of up to 322500 k-points per $\text{\AA}^{-2}$ and BZ sizes of up to $11.34\text{\AA}^{-1}\times2.65\text{\AA}^{-1}$. 
Time integration is performed using a fourth order Runge-Kutta solver with time steps of at least $0.005$ fs.
We show in Fig.~\ref{fig01_convergence} and \ref{fig023_convergence}, that increasing the BZ size, k-point density and decreasing the time step of the integration does not change the values of our calculations
We employ an electric field, that is linearly polarized in x-direction. 
Unless otherwise stated, we employ a gapless Dirac cone with an equilibrium band occupation of a Fermi-Dirac distribution with an Fermi-Level of $0.2$ eV at zero temperature. 
The offdiagonal dephasing rate $\gamma$ for all plots besides Fig.~(2) is chosen to be $(10\,\text{fs})^{-1}$~\cite{Floss2018}.


\bibliography{Literature}
\bibliographystyle{apsrev4-2.bst}